\documentclass[12pt]{article}
\usepackage[a4paper]{geometry}
\usepackage{lineno}
\usepackage{amssymb}
\usepackage{amsmath}
\usepackage{amsthm}
\usepackage{epsfig}
\usepackage{graphicx}
\usepackage{graphics}
\usepackage{float}
\usepackage{subfigure}
\usepackage{multirow}
\usepackage{color}
\usepackage{lineno}
\usepackage{fullpage}
\usepackage[normalem]{ulem} 
\usepackage{makeidx}
\usepackage{xspace}
\usepackage{wrapfig}
\usepackage{multicol}
\usepackage[affil-it]{authblk}
\usepackage{setspace}

\begin{document}
\onehalfspacing 

\title{Green's functions and method of images: an interdisciplinary topic usually cast aside in physics textbooks}

\author{Glauco Cohen Ferreira Pantoja%
  \thanks{Electronic address: \texttt{glaucopantoja@hotmail.com};}}
\affil{Instituto de Ciências da Educação,\\ Universidade Federal do Oeste do Pará, Brazil}

\author{Walace S. Elias%
  \thanks{Electronic address: \texttt{walace.elias@ufra.edu.br}}}
\affil{Instituto Ciberespacial,\\ Universidade Federal Rural da Amazônia, Brazil}

\date{}

\maketitle

\begin{abstract}
In the present work we discuss how to address the solution of electrostatic problems, in professional cycle, using Green's functions and the Poisson's equation. For this, it was considered the structural role that mathematics, specially Green's function, have in physical thought presented in the method of images. By using this procedure and discussing the historical construction of Green's problem, it was possible to verify its relation with the method of images as an interdisciplinary approach which is not developed in didactic physics textbooks. The possibility of analyzing Green's function as the result of two tasks, namely, the reduction of a continuous charge distribution to the one due to a point charge and the solution of the problem as the superposition of potentials due to sets of point charges continuously distributed represented by the integration of the Green's function over the electric charge density, is our account for teaching implication that shows, at the same time, epistemological ruptures and continuities to teaching-learning processes.
\end{abstract}

\maketitle


\section{Introduction}

One of the usual learning tasks on electrostatics consists in obtaining the electric field (or potential) generated by a charge distribution in certain region of the space. It is possible to solve this problem by direct integration over the charge distribution or tackling the Poisson's equation subjected to a set of boundary conditions imposed on the field.

Solving non-homogeneous differential equations u\-sing Green's Functions is one of the most powerful forms of describing the solution for a problem of this kind.  However, a great number of classical books on electrodynamics do not explore Poisson's Equation solutions using this method. Instead, the common approach uses the method of images, a very interesting way to solve the problem, once it requires a deep physical interpretation. 

The method of images can be used when we are trying to obtain the electrostatic field generated by charge distributions near a conductive surface.  This procedure takes into account the symmetry of the problem by adding an image charge outside the region of interest. From this new arrangement, it's possible to reconstruct the same boundary conditions of the initial problem without these image charges. 

However, it seems at first that this procedure is barely or not related to the resolution of Poisson's (or Laplace) equation. The method of images, on the contrary,  is completely compatible with the more general procedure of solving Poisson's equation via Green's functions. The former is a practical and conceptually elegant mathematical tool, even though it is not general as it is the latter. In addition, it assumes the existence of virtual image charges in regions where the solution is not valid, what seems somewhat artificial for introducing in teaching and learning processes.

Author and Colleague \cite{Author} developed a classification of tasks on electrostatics in which are presented four primary classes of situations addressed to: \textbf{(i)} calculation of electrostatic fields; \textbf{(ii)} symbolic representation of the electrostatic field; \textbf{(iii)} analogical representation of the electrostatic field; \textbf{(iv)} description of electrostatic interactions. 

All these primary classes include four or five secondary classes of situations whose classification is based on the objects, variables and undetermined quantities or qualities  presented by the problem. According to authors in case (i) there are five classes of situations. Among them the most complicated/complex is the one related to calculation of electrostatic fields (or potentials) due to unknown charge distributions, which include, for instance, conductors in electrostatic equilibrium. 

Their arguments rely upon two main epistemological reasons related to the necessary thought operations for mastering these problems: advanced mathematical techniques for problem-solving and conceptual deepness demanded for physical interpretation of charge redistribution.

The relation among these concepts has been barely explored and, moreover, the calculations made by the methods of images are often restricted to the case of point charges, with exception to Reitz, Milford and Christy \cite{Reitz} that include the problem of linear images, cases in which it is possible to calculate the potencial due to very large, electrically charged, wires placed in regions containing conductors.

Although Machado \cite{Machado} and Jackson \cite{Jackson} solve the problem of a discharged and grounded sphere in front of a point charge through the method of Green's function, more complicated situations involving electric charge distributions are rarely discussed using this procedure. 

For instance, Panofsky and Phillips \cite{Panofsky} approach the general problem of Green's function by discussing general, mathematical and physical features, however they do not elaborate the discussion for specific problems. In this work we present a discussion on how to approach electrostatics in the professional cycle, from the point of view of solving Green's functions for Poisson's equation, can be articulated to the method of images in an interdisciplinary approach. Our framework takes into account the structural role that Mathematics (Green's functions) have in Physical thought (method of images).

The structure of this paper is presented as follows. Section \ref{Frame} presents a brief discussion about the structural role of mathematics in physical thought. In Section \ref{Met}, we present the problem of Green's functions from a historical point of view and the complete mathematical formulation of the solution of the Poisson's equation, considering three-dimensional and two-di\-men\-si\-o\-nal cases.  Section \ref{Examples} presents a set of electrostatic problems, whose solution was obtained by Green's function, to verify the relation with method of images. Final remarks are made in the Section \ref{Conclusion}.

\section{Mathematics structures Physical thought} \label{Frame}

The relation between Mathematics and Physics is not just historical, but also epistemological. The junction among Physics, Astronomy and Mathematics in the Copernican Revolution fully stresses this fact. Expressing physical ideas in mathematical terms, on the other hand, is much more than a predictive tool, because it envolves structuring physical thought in function of mathematic enunciations. It is not necessary to defend the role of Mathematics in Physics, because it is blatantly obvious. However, it is fundamental to discuss which role of Mathematics is developed in teaching and learning these disciplines.

Karam \cite{Karam} and Rebello et al.\cite{Rebello} state that the results of studies on transference from Mathematics to Physics are strikingly clear about the hindrances faced by the students in this task, once using the former in the latter envolves more than a simple correspondence relation between two distinct conceptual domains. In other words, that means this association is very different from the rote use of formulas.

Thus, approaching the role of Mathematics in Physics\sout{,} requires differentiating its technical role (tool-like) and its structural role (reason-like). The first one can be assumed when it is used in the second one. In table \ref{tabela1} some characteristics of the technical dimension concerning the role of mathematics are pointed (extracted from Karam \cite{Karam}).

Therefore, Karam \cite{Karam} states that the \textit{technical role} of Ma\-thematics is associated with calculations developed in a disconnected way from physical problems (e.g plug-and-chug), while its counterpart, the structural one, is related to the use of Mathematics to reason about the physical world, that is, to establish reference to it. Although the first is important for mastering the second one, the technical domain is not sufficient to lead students to the structural level \cite{Karam}. The author highlights it is impossible to detach conceptual understanding and mathematical structures use, and points some important characteristics of this feature, which we present in the table \ref{tabela2}.

We then seek to discuss the structural role of Green's function in Physics by explaining its relation with the Method of Images. It is possible to do it by modelling and comprehending problems containing known and unknown charge distributions.

\begin{table}[ht!]
\begin{tabular}{|p{8.4cm}|}
\hline
\hspace{30mm} \textbf{Technical} \\
\hline
1) Blind use an equation to solve quantitative problems;\\ 
2) Focus on mechanic or algorithmic manipulations;\\
3) Utilization of arguments of authority;\\
4) Rote memorization of equations and rules;\\
5) Fragmented knowledge;\\
6) Identification of superficial similarities between equations;\\
7) Mathematics conceived as calculation tool;\\
8) Mathematics seen as language used to represent and communicate;\\
\hline
\end{tabular}
\caption{Technical dimension concerning the role of mathematics in physics.}
\label{tabela1}
\end{table}
\begin{center}
\begin{table}[ht!]
\begin{tabular}{|p{8.4cm}|}
\hline
\hspace{30mm} \textbf{Structural} \\
\hline
1) Derive an equation from physical principles using logical reasoning;\\ 
2) Focus on physical interpretations or consequences;\\
3) Justify the use of specific mathematical structures to model physical phenomena;\\
4) Structured knowledge: connect apparently dif\-fe\-rent physical assumptions through logic;\\
5) Recognition of profound analogies and common ma\-the\-ma\-ti\-cal structures behind different physical phe\-no\-me\-na;\\
6) Mathematics conceived as reasoning instrument;\\
7) Mathematics seen as essential to define physical concepts and structure physical thought;\\
\hline
\end{tabular}
\caption{Structural dimension concerning the role of mathematics in physics.}
\label{tabela2}
\end{table}
\end{center}

In summary, the main difference between the two perspectives underlies the fact that the technical view uses mathematics merely as a tool to solve physics problems, while the structure view uses it to model physics problems and to relate the objects, variables and undetermined quantities in order to find a meaningful solution to it, that is, to structure quantitative thinking processes.

\section{Green's Functions and Poisson's equation} \label{Met}

In this section, the problem of Green's function is presented from a historical point of view and the apparent contradiction in the fact that differential operators applied in Green's Functions are expressed in terms of the Dirac Delta "function" \cite{Cannell} is discussed. 

\subsection{A brief history of Green's Functions}

How could Green be alive between XVIII and XIX centuries to write his formulation in XX century notation? The reason is: he did not do that. 

George Green was born on July 14th 1793, No\-ttin\-gham, England and died on May 31st 1841, in the same town. It was one of the biggest exponents in Mathematical-Physics of the region, being the first to introduce the concept of potential and the method of Green's functions, largely used until the present days in many fields on Physics. However, it seems he has been forgotten for a while, what would imply posthumous recognition for his work, due to his popularisation in works of Wi\-lli\-am Thom\-son, known as Lord Kelvin \cite{Cannell}.

Nonetheless, if his work was so important both for Mathematics and Physics, why it remained obscure in history? Cannell \cite{Cannell} enumerates factores like: his premature death, at the age 47; the fact of going to Cambridge to study lately and then returning to Nottingham, without establishing personally in the former city; his graduation in math in a relatively advanced age; the development of abstract works for the period he lived, without drawing attention of the scientific community, which was worried with practical questions at that time; the advanced nature of his work, barely understood for much scientists of the \textit{époque}.

Electromagnetism was not a commonplace subject at Green's time, it became so solely after Kelvin and Faraday. Green knew, however, the works of Laplace, Legendre and Lacroix and had access to a translation to english of the \textit{Mècanique Celeste} due to Pierre Laplace, made by John Toplin, his tutor in \textit{Nottingham Free Grammar School}, a Leibnizian (what explains his preference for "d-ism" instead of the Newtonian "dot-ism"). Green also deeply knew the work of Poisson on Magnetism, probably accessed by attending to the \textit{Nottingham Subscription Library}. The mathematician was interested, in his essay on electricity and magnetism (1828), in inverse-type problem related to the electric potential (physical quantity named after Green), namely, "knowing the potential how can we determine the electric fluid (electric charge) density in a ground conductor of any form?''\cite{Cannell}. The former solution to Poisson's equation was given in modern notation by
\begin{eqnarray}
\nabla^2\varphi &=& -\frac{\rho}{\epsilon_0}
\\
\varphi &=& \frac{1}{4\pi\epsilon_0}\int_V\frac{\rho(\vec{r}^{\prime})}{\vert\vec{r}-\vec{r}'\vert}dV'.
\end{eqnarray}

In other words, Green was interested in determining the charge distribution from operations on the potential function, whose negative gradient would result in the force on an unit charge exerted on this conductor (nowadays, it is interpreted as the electric field). 

Green developed a work in mathematical analysis and constructed what we know by Green's theorem\cite{Cannell}. After using his theorem, Green investigated what happens in the neighborhood of a point charge located in $\vec{r}=\vec{r'}$ (modern notation) by evaluating the limit of the solution \textcolor{blue}{when} $\vec{r}\rightarrow\vec{r}'$, namely, when the evaluation point tends to the position of the point charge. He then carried out to the following function (modern notation)\footnote{Once we changed the original Poisson's equation for the international system, the original Green function was $G(\vec{r},\vec{r}')=\frac{1}{\vert\vec{r}-\vec{r}'\vert}$}:
\begin{equation}
G(\vec{r},\vec{r}')=\frac{1}{4\pi\vert\vec{r}-\vec{r}'\vert}.
\end{equation}

Green also applied his solution and succeeded in finding a formula relating the unknown surface charge density in a condutor with the known potential in its surface; his solution is likewise discontinuous, what is physically feasible, once electric charges (or fluids, at that time), were known to stay concentrated in the conductor's surface. The mathematician checked if the function satisfied Laplace's equation outside the source and considered the Green function as a response to an unitary impulse \cite{Cannell}, exactly as is done nowadays.

In 1930, 102 years later, Paul Dirac introduced his famous "delta functions'' without proper mathematical rigor, although with a significant practical value. In modern notation the differential equation satisfied by Green's functions are presented in function of these ``improper functions'', as Dirac called them. 

Nevertheless, the formalisation of such mathematical elements just became possible after the work of Laurent Schwartz on the theory of distributions in the 50's , in which he describes the "delta functions'' as limits of a sequence, \textit{id est}, a distribution \cite{Schwartz}. This is the reason why, in our calculus, we use modern notation to find Green's functions. We can clearly see how Green was ahead of his time.

It can be seen that in the history of construction of Green's function, an interdisciplinary relation between mathematics and physics, based in mathematics structuring the physical though, allowed the construction of a fruitful research program for both disciplines. Green basically built a mathematical idea departing from reasoning with physical quantities. Schwartz, on the other hand, assembled a formal structure that permits analysing physics knowledge in a more organised and formal structure.
\subsection{Poisson's Equation}

Electric charges are held stationary by other forces than the ones of electric origin, such as molecular binding forces. Since charges are stationary, no electric currents and, thus, no magnetic fields are presented ($\vec{B}=0$). For a stationary electric charge distribution, described by $\rho(\vec{r})$, the associated electrostatic field satisfies the following set of differential equations, 
\begin{eqnarray}
\label{divE}
\nabla\cdot\vec{E}=\frac{\rho}{\epsilon_0},
\\
\label{rotE}
\nabla\times\vec{E}=0.
\end{eqnarray}
\noindent Accordingly to the Helmholtz's theorem \cite{Griffiths}, if both divergence and curl of sufficiently smooth, rapidly decaying, vector fields are known, the problem can be solved. For the electrostatic field, the solution for $\vec{E}$ can be written as the gradient of a scalar function $\varphi(\vec{r})$, since it is irrotational:

\begin{equation}
\vec{E}=-\nabla\varphi,
\label{potencial}
\end{equation}
where $\varphi(r)$ is well-known as the electrostatic potential. Replacing \eqref{potencial} in the equation \eqref{divE}, leads to the Poisson's equation:
\begin{equation}
\nabla^2\varphi=-\frac{\rho}{\epsilon_0},
\label{6}
\end{equation}
and when regions without electric charge distribution are considered, $\rho = 0$, equation (\ref{6}) becomes
\begin{equation}
\nabla^{2}\varphi = 0,
\label{laplace_equation}
\end{equation}
which is known as Laplace's equation.

In electrostatics problems the solution will be unique if boundary conditions are imposed on the potential $\varphi(\vec{r})$ (on the electrostatic field $\vec{E}(\vec{r})$) in some point of the space, accordingly to the Uniqueness theorem \cite{Jackson}. If boundary conditions are imposed on $\varphi(\vec{r})$, these are known as Dirichlet boundary conditions. However, when boundary conditions are applied to $\vec{E}(\vec{r})$, they are denoted as Neumann bo\-un\-da\-ry conditions \cite{Jackson}. Another possibility is to apply mixed boundary conditions, both on $\varphi(\vec{r})$ and $\vec{E}(\vec{r})$. In this case, we call Robbin's boundary conditions.

\subsection{Green's functions}

In general, solving the scalar differential equation \eqref{6} for $\varphi$ is easier than solving vector differential equations \eqref{divE} and \eqref{rotE}. We can apply Green's function in \eqref{6}, obtaining the n-dimensional equation below,
\begin{equation}
\nabla^2G(\vec{r},\vec{r}^{\prime})=-\delta^{(n)}(\vec{r}-\vec{r}^{\prime}).
\label{Green_function_Poisson}
\end{equation}

Considering the Green's identities \cite{Bassalo, Arfken}, it is possible to obtain an expression for electrostatic potential
\begin{eqnarray}
\label{green1identity}
\nabla\cdot(\varphi\nabla G)=\varphi\nabla\cdot(\nabla G)+\nabla\varphi\cdot\nabla G,
\\
\label{green2identity}
\nabla\cdot(G\nabla\varphi)= G\nabla\cdot(\nabla\varphi)+\nabla\varphi\cdot\nabla G,
\end{eqnarray}
which leads to
\begin{equation}
\int \left(\varphi\nabla^2G-G\nabla^2\varphi\right)dV=\oint\left(\varphi\nabla G-G\nabla\varphi\right)\cdot \hat{n}dS.
\label{8}
\end{equation}
Then, using \eqref{6} and Green's Identity \eqref{green1identity} $-$ \eqref{green2identity} in equation \eqref{8}, one can obtain
\begin{equation}
\varphi(\vec{r})=\frac{1}{\epsilon_0}\int G\rho dV'+\oint G\frac{\partial \varphi}{\partial n}dS'-\oint \varphi\frac{\partial G}{\partial n}dS',
\label{greengeneralsolution}
\end{equation}
which is the general solution for an electrostatic potential and, consequently, for the electric field. The second and third terms in equation \eqref{greengeneralsolution} are associated with the choice of the boundary conditions to which electric charge density is subject. Once boundary conditions are defined, equation \eqref{greengeneralsolution} will have an unique and well-defined solution accordingly to the uniqueness theorem. 

In a great number of physical problems that include conductors in electrostatic equilibrium and zero potential, it is adequate to apply Dirichlet's boundary conditions on both Potential and Green's Function, which implies
\begin{equation}
\varphi(\vec{r})=\frac{1}{\epsilon_0}\int G\rho dV'.
\label{general_potencial_green}
\end{equation}

For this kind of problem it is always possible to add into the Green's function a solution to Laplace's equation, denoted by $G_{\mathsf{L}}(\vec{r},\vec{r}'$), which satisfies physical and mathematical boundary conditions. Therefore, the full Green's function will be written as
\begin{equation}
G(\vec{r},\vec{r}^{\prime})=G_{\mathsf{D}}(\vec{r},\vec{r}^{\prime})+ G_{\mathsf{L}}(\vec{r},\vec{r}')\,,
\end{equation}
where  $G_{\mathsf{D}}(\vec{r},\vec{r}^{\prime})$ de\-pends ex\-clu\-si\-ve\-ly the di\-men\-si\-ons of Laplacian operator, whe\-reas $G_{L}(\vec{r},\vec{r}^{\prime})$ depends on chosen boundary conditions. In the next section, we shall determine the expression for $G_{\mathsf{D}}$ in 3-dimensional and 2-dimen\-sional cases for Laplacian operator.

\subsubsection{Green's function for Poisson's Equation}

For didactical reasons, we determine the expression for the Green's function of Poisson's equation considering the 3-dimensional and the 2-dimensional cases in two different sections.

\subsubsection*{Three-dimensional case}
The analytical expression for Green's function in three dimensions will be determined. It is necessary to apply a Fourier Transform, leading Green's function to $k-$space. 

The Fourier transform and its inverse for Green's function $G_{\mathsf{D}}$ are presented below, respectively
\begin{eqnarray}
\label{Fourier_G}
G^{\mathsf{D}}_{k} \equiv G_{\mathsf{D}}(\vec{k},\vec{r}^{\prime})=\int_{-\infty}^{\infty}G_{\mathsf{D}}(\vec{r},\vec{r}^{\prime})e^{i\vec{k}\cdot\vec{r}}d^3r,
\\
G_{\mathsf{D}}(\vec{r},\vec{r}^{\prime})=\frac{1}{(2\pi)^3}\int_{-\infty}^{\infty}G^{\mathsf{D}}_{k}\,e^{-i\vec{k}\cdot\vec{r}}d^{3}k.
\end{eqnarray}

Applying the Fourier transform\eqref{Fourier_G} on equation \eqref{Green_function_Poisson}, for $n=3$, integrating by parts we obtain the following expression
\begin{equation}
G^{\mathsf{D}}_{k}=\frac{e^{i\vec{k}\cdot\vec{r}^{\prime}}}{\left(k_x^2+k_y^2+k_z^2\right)},
\label{eq18}
\end{equation}
which represents Green's function in $k-$space. Applying the inverse Fourier transform in \eqref{eq18}, we will recover the expression for $G_{\mathsf{D}}$ in coordinates space,
\begin{equation}
G_{\mathsf{D}}(\vec{r},\vec{r}')=\frac{1}{(2\pi)^3}\int_{-\infty}^{\infty}\frac{e^{-i\vec{k}\cdot\vec{R}}}{\left(k_x^2+k_y^2+k_z^2\right)}d^3k,
\label{eq19}
\end{equation}
with $\vec{R}=\vec{r}-\vec{r}^{\prime}$. The integral in equation \eqref{eq19} becomes simpler by an adequate change of variables. Once the integrand does not depend on variable $\phi$, integration results in numerical factor equals to $2\pi$. Rewriting the exponent in equation \eqref{eq19} as $\vec{k}\vec{R}=kRcos\theta$, we integrate over variable $\theta$ to find the following result
\begin{equation}
G_{\mathsf{D}}(\vec{r},\vec{r}')=\frac{1}{2\pi^2 R}\int_0^{\infty}\frac{\sin(kR)}{k}dk.
\label{eq20}
\end{equation}

The integral in \eqref{eq20} can be taken to the complex plane, with part of it being an integral along the real axis and the other one along a contour $\Gamma$ extending to infinity\cite{Arfken}

\begin{equation}
\oint \frac{\sin(z)}{z}dz=\int_{-\infty}^{\infty}\frac{\sin(z)}{z}dz
=\frac{1}{2i}\oint\limits_{\Gamma} \frac{e^{iz}-e^{-iz}}{z}dz,
\end{equation}

We find the following result 

\begin{equation}
\int_{-\infty}^{\infty}\frac{\sin(z)}{z}dz=\frac{2\pi i}{2i}e^0=\pi.
\end{equation} 

and we can find the the Green's function for Laplace's equation for the three-dimensional case \eqref{eq20}, 

\begin{equation}
G_{\mathsf{D}}(\vec{r},\vec{r}^{\prime})=\frac{1}{4\pi\vert{\vec{r}-\vec{r}'}\vert}.
\label{green_3d}
\end{equation}

We will discuss the two-dimensional Laplacian operator case in sequence.

\subsubsection*{Two-dimensional case}

Evoking the the Green Function for the two-di\-men\-si\-o\-nal Poisson's equation 
\eqref{Green_function_Poisson}, for $n=2$,

\begin{equation}
\nabla^2G_{\mathsf{D}}(\vec{r},\vec{r}^{\prime})=-\delta^2(\vec{r}-\vec{r}^{\prime}),
\label{LFG}
\end{equation}

\noindent\textcolor{blue}{and} enunciating both Fourier direct and inverse transforms
\begin{eqnarray}
G^{\mathsf{D}}_k=G_{\mathsf{D}}(\vec{k},\vec{r}^{\prime})=\int_{-\infty}^{\infty}G_{\mathsf{D}}(\vec{r},\vec{r}^{\prime})e^{i\vec{k}\cdot\vec{r}}d^2r,
\\
G_{\mathsf{D}}(\vec{r},\vec{r}^{\prime})=\frac{1}{(2\pi)^2}\int_{-\infty}^{\infty}G^{\mathsf{D}}_{k}e^{-i\vec{k}\cdot\vec{r}}d^{2}k,
\end{eqnarray}
we can follow the similar procedure in three-di\-men\-si\-o\-nal case and obtain
\begin{eqnarray}
G^{\mathsf{D}}_{k}&=&\frac{e^{i\vec{k}\cdot\vec{r}'}}{k_x^2+k_y^2},
\label{eq27}
\\
G_{\mathsf{D}}(\vec{r},\vec{r}^{\prime})&=&\frac{1}{(2\pi)^2}\int_{-\infty}^{\infty}\frac{e^{i\vec{k}\cdot\vec{r}'}e^{-i\vec{k}\cdot\vec{r}}}{k_x^2+k_y^2}d^2k.
\label{eq28}
\end{eqnarray}

To solve the integration in equation \eqref{eq28}, we can change variables to polar coordinates and apply the scalar product
\begin{equation}
G_{\mathsf{D}}(\vec{r},\vec{r}^{\prime})=\frac{1}{(2\pi)^2}\int_{0}^{\infty}\int_0^{2\pi}\frac{e^{ikR\,\cos\theta}}{k}dkd\theta,
\end{equation}
recognizing the integral in $\theta$ as $2\pi J_0(kR)$, where $J_0(kR)$ is the zero-order Bessel function
\begin{equation}
\begin{split}
G_{\mathsf{D}}(\vec{r},\vec{r}^{\prime})&=\frac{1}{(2\pi)}\int_{0}^{\infty}\frac{J_0(kR)}{k}dk 
\\
&= -\frac{1}{2\pi}\ln \vert\vec{r}-\vec{r}'\vert
\end{split}
\label{eq30}
\end{equation}
where  the equation \eqref{eq30} represents the Green's function for two-dimensional Laplace's case.

\section{Solving electrostatic problem by Green's Function}
\label{Examples}

In the previous section we presented the expressions of Green's functions in 3-dimensional and 2-dimensional cases. Now, we will drive our attention to obtain the solution for a set of problems using the Green's function and verifying the relation with the method of images.
\subsection{Point Charge placed near a Grounded Infinite Plane Conductor}

Let's consider a point electric charge, $q$, placed a distance $d$ along the $z$ axis of an infinite thin grounded plate along the $xy$ plane.What is electrical potential produced in a region $z>0$ in space? 

Green's function for the three-dimension problem, in this case, admits a solution for Laplace's equation adjusted to Dirichlet boundary conditions for both Green Function and Electric Potential. 

The ground conductor is mathematically structured as having null electric potential over its surface at $z=0$. Meanwhile, Green's Function reduces the problem of a continuous, and in this case unknown, distribution to the one of a point charge, exactly what the method of images proposes. 

Considering the equation \eqref{green_3d}, the full Green's function to this problem will be
\begin{equation}
\begin{split}
G(\vec{r},\vec{r}^{\prime})&=\frac{1}{4\pi\sqrt{(x-x')^2+(y-y')^2+(z-z')^2}}+
\\
&+G_{\mathsf{L}}(\vec{r},\vec{r}').
\end{split}
\end{equation}
On the boundary $\vec{S}=(x,y,z=0)$, for every point located on the plate, Green's function equals zero
\begin{equation}
G(S,\vec{r}^{\prime})=0.
\label{eq34}
\end{equation}

Thus, it is possible to see by inspection from equation \eqref{eq34}, that the function $G_{\mathsf{L}}$ must have the following expression
\begin{equation}
G_{\mathsf{L}}(\vec{r},\vec{r}')=
-\frac{1}{4\pi\sqrt{(x-x')^2+(y-y')^2+(z+z')^2}},
\end{equation}
to ensure the condition given by equation \eqref{eq34} will be valid. Therefore, insofar the only known electric charge is a point one, it can be modeled by a Dirac Delta function charge density, whose infinity point is located at point $(0,0,d)$.

In other words, the eletric charge density can be write as $\rho(x,y,z)=q\,\delta(z-d)\,\delta(y-0)\,\delta(x-0)$ in such a way that integrating over the volume in equation \eqref{general_potencial_green} leads to:
\begin{equation}
\begin{split}
\varphi(x,y,z)&=\frac{q}{4\pi\epsilon_0\sqrt{(x)^2+(y)^2+(z-d)^2}}+
\\
&-\frac{q}{4\pi\epsilon_0\sqrt{(x)^2+(y)^2+(z+d)^2}}.
\end{split}
\label{potencial_cargas}
\end{equation}

\begin{figure}[htp]
  \centering
   \includegraphics[width=\linewidth]{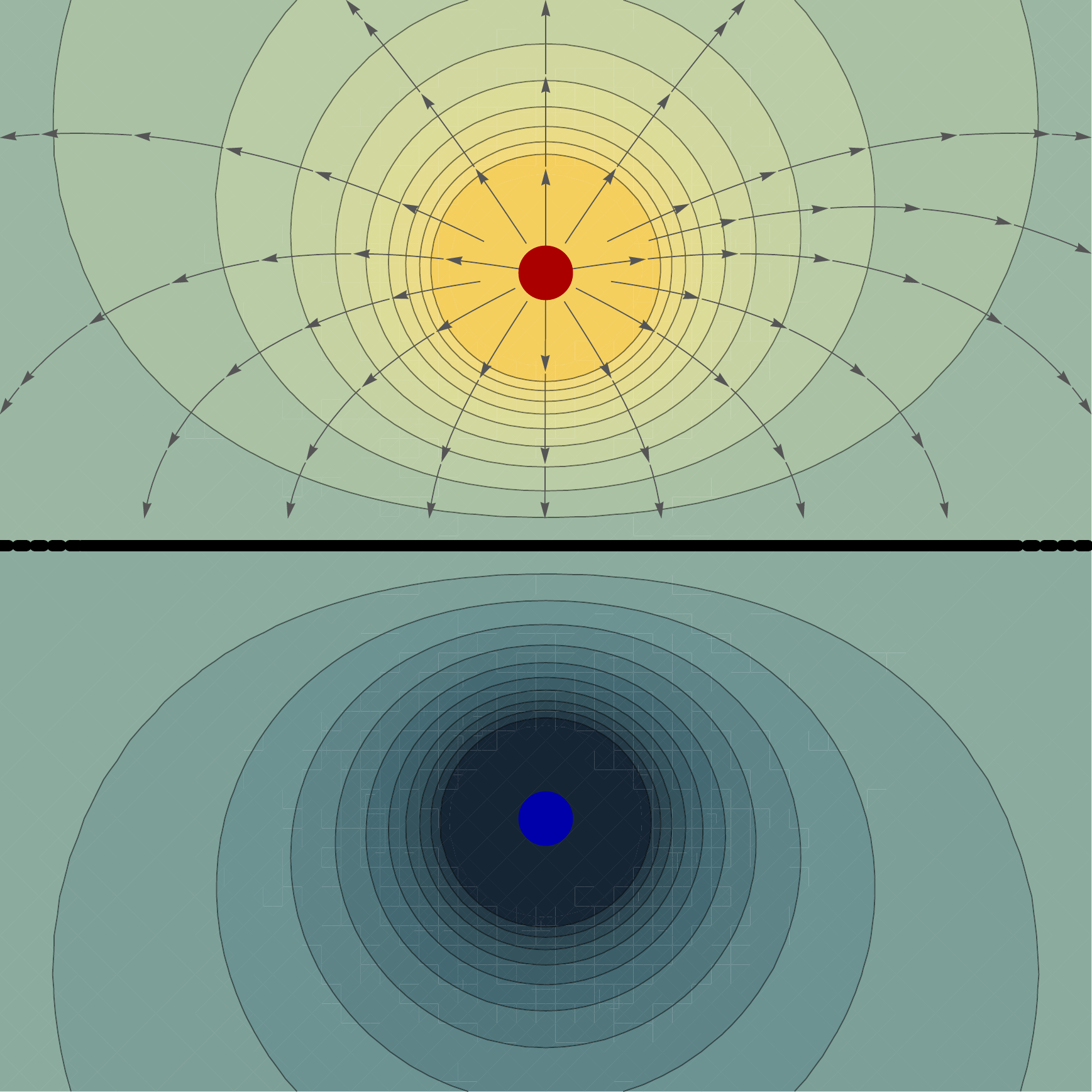}
	    \caption{Contour lines for the electrostatic potential \eqref{potencial_cargas}. We can verify where we must place an image charge (blue) in a way that maintains a null potential over the grounded plane conductor (black thick line)\textcolor{red}{.}} 
	        \label{curvas_nivel_cargas}
\end{figure}

The equation \eqref{potencial_cargas} is a solution for Laplace's Equation for $z>0$, except for $z=d$ where it diverges. Besides, the solution satisfy the imposed boundary conditions. The result is the same obtained by the method of images as seen in figure \ref{curvas_nivel_cargas}. 

It is important to highlight the symmetry of Green's Function in the inversion between point and source locations. Structuring Physical thought (withdraw the plane by a point charge) in such a way to make the Electrical Potential to vanish in that surface is a matter related to the mathematical point of view. 

\subsubsection*{Electric Field}

After determining the electric potential, it is possible to find the associated electric field. Therefore, using the equation \eqref{potencial}, we find
\begin{equation}
\vec{E}=\frac{q}{4\pi \epsilon_{0}}\left(E_{x}\hat{i}+E_{y}\hat{j}+E_{z}\hat{k}\right),
\end{equation}

where 

\begin{equation}
E_x=\left[\frac{x}{\left((d-z)^2+r^2\right)^{3/2}}-\frac{x}{\left((d+z)^2+r^2\right)^{3/2}}\right],
\end{equation}
and
\begin{equation}
E_y=\left[\frac{y}{\left((d-z)^2+r^2\right)^{3/2}}-\frac{y}{\left((d+z)^2+r^2\right)^{3/2}}\right],
\end{equation}
and
\begin{equation}
E_z=\left[\frac{d-z}{\left((d-z)^2+r^2\right)^{3/2}}-\frac{d+z}{\left((d+z)^2+r^2\right)^{3/2}}\right],
\end{equation}

with $r^2=x^2+y^2$, represents the components of three dimensional electric field.

\subsection{Infinite charged wire placed near a Grounded Infinite Plane Conductor}

Consider an infinite charged wire placed at distance $d$ along the $x$ axis, near a grounded infinite plane conductor. It is possible to use the Green function in two-dimensional case \eqref{eq30}, to adjust a solution to Laplace's equation $G_{\mathsf{L}}$. The potential is zero on the char\-ged pla\-ne, which leads to the full Green's function

\begin{equation}
G(x,x',y,y')=-\frac{1}{2\pi}\ln\sqrt{\frac{(x-x')^2+(y-y')^2}{(x+x')^2+(y-y')^2}},
\label{fio_green}
\end{equation}

where we considered the Dirichlet for the determination of the Green's function  $G_{L}$. The equation \eqref{fio_green} represents the same result obtained by the method of images. 

To obtain the electrostatic potential for this case, we must integrate the Green function over volume in equation \eqref{general_potencial_green}. Considering the charge density function given by $\rho=\lambda\delta(x-d)\delta(y-0)$, one obtains

\begin{equation}
\varphi(x,y)=-\frac{\lambda}{2\pi\epsilon_0}\ln\sqrt{\frac{(x-d)^2+(y)^2}{(x+d)^2+(y)^2}}\,\,.
\label{fio_potencial}
\end{equation}

From equation \eqref{fio_potencial} it is possible to study the equipotential surfaces\textcolor{blue}{.} If the argument of the logarithm function is a constant
\begin{equation}
\frac{(x-d)^2+y^2}{(x+d)^2+y^2}=m,
\label{eq43}
\end{equation}
thus represents a circumference with equation

\begin{equation}
\left[x-\left(d\frac{(1+m^2)}{(1-m^2)}\right) \right]^2+y^2=\left(\frac{2md}{1-m^2}\right)^2.
\end{equation}
\begin{figure}[htp]
  \centering
\includegraphics[width=\linewidth]{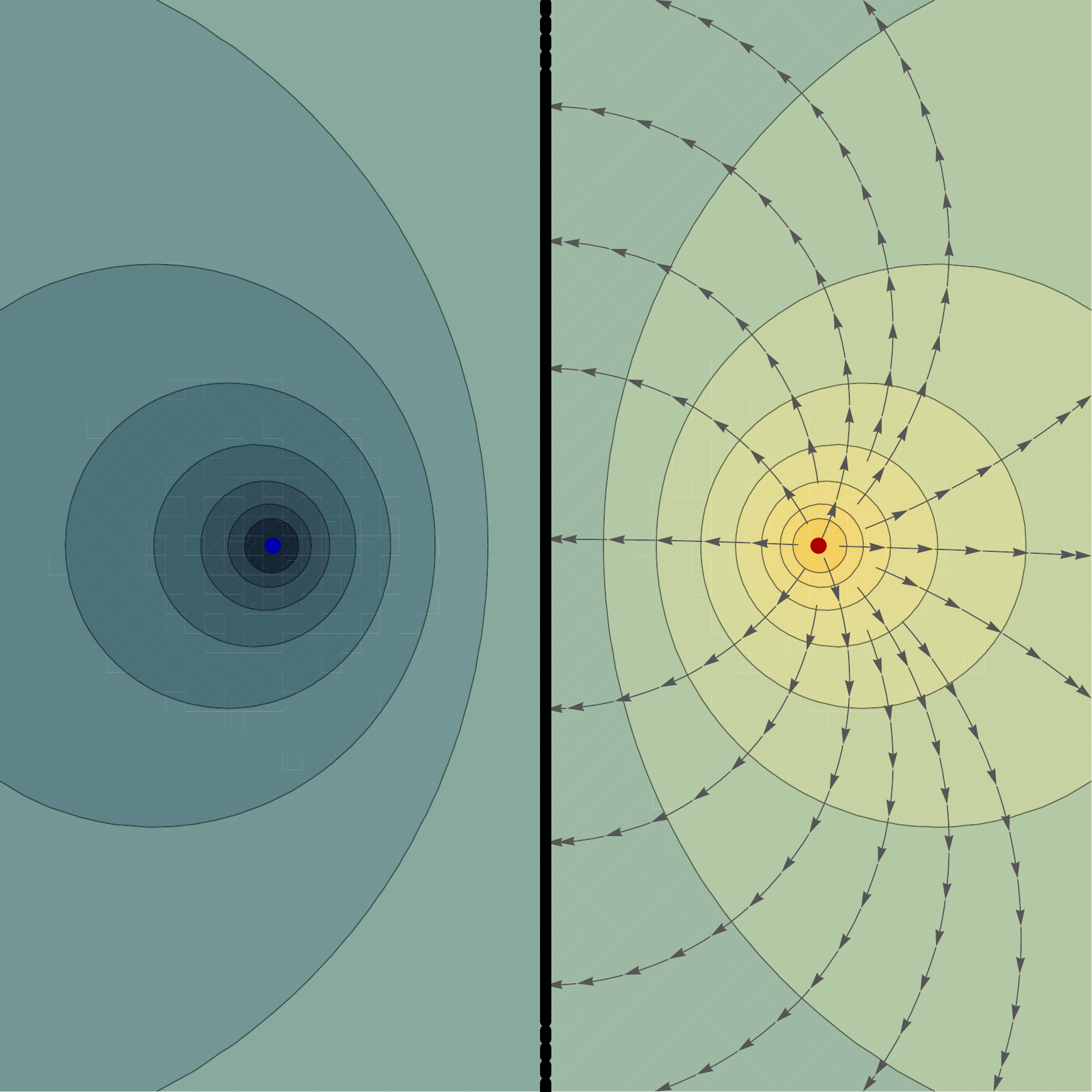}
    \caption{Circular equipotential surfaces for the electric potential due to a charged wire next to an infinite plane}
       \label{surface_level_potential}
\end{figure}

For the case $m=1$ in equation \eqref{eq43}, the radius of the circumference will be infinite, which represents a plane. As long as the solution fits $\varphi(S)=0$ and $x_0=\infty$, the equipotentials are on the plane and at infinity. 
Considering the cases with $m<1$, the equipotentials surfaces represent circles of radii $r=\frac{2md}{1-m^2}$ centered at point $x_0=d\frac{(1+m^2)}{(1-m^2)}$, as presented in figure \ref{surface_level_potential}.

\subsubsection*{Electric Field}

From the electrostatic potential \eqref{fio_potencial}, the resultant electric field can be found using equation \eqref{potencial},  given by the following expression
\begin{equation}
\vec{E}(\vec{r})=E_{x}\hat{i}+E_{y}\hat{j},
\label{campo_eletrico_fio}
\end{equation}
where
\begin{equation}
E_{x}= \frac{\lambda\,d}{\pi \epsilon_{0}} \left\lbrace\frac{d^{2}-x^{2}+y^{2}}{[(d-x)^{2}+y^{2}][(d+x)^{2}+y^{2}]}\right\rbrace,
\end{equation}
and
\begin{equation}
E_{y}=\frac{\lambda\,d}{\pi \epsilon_{0}}\left\lbrace\frac{2\,x\,y}{[(d-x)^{2}+y^{2}][(d+x)^{2}+y^{2}]}\right\rbrace,
\end{equation}
represents the two cartesian coordinates of electric field.
\subsection{Point Charge placed near a Grounded Spherical Conductor}

\begin{figure}[htp]
  \centering
\includegraphics[width=\linewidth]{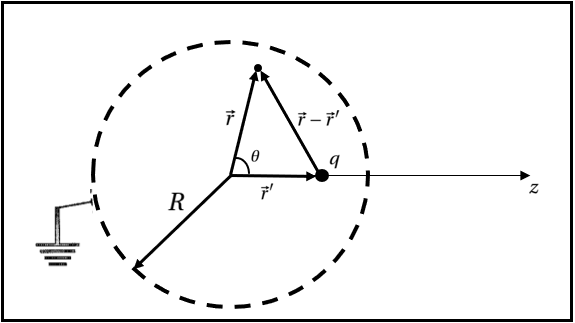}
    \caption{Diagram illustrating the Laplace's equation for a sphere of radius R, with a point charge located at $\vec{r}^{\prime}$. }
    \label{esfera}
\end{figure}

We shall solve the classical problem of finding the potential inside a grounded sphere of radius $R$, centered at the origin, due to a point charge inside the sphere at position $\vec{r}^{\prime}$, as showed at figure \ref{esfera}. 

The full Green Function for this problem is given by
\begin{equation}
\begin{split}
G(r,\theta,r',\theta')&=\frac{1}{4\pi}\frac{1}{\sqrt{r^2+r'^2-2rr'\cos(\theta-\theta')}}
\\
&+G_{\mathsf{L}}(r,\theta,r',\theta'),
\end{split}
\label{full_green_function}
\end{equation}
where we already considered the Green's function \eqref{green_3d} and the spherical symmetry of the problem to write the distance $|\vec{r}-\vec{r}^{\prime}|$.

In the similar way, it is necessary to add $G_{\mathsf{L}}$ into the full Green's function \eqref{full_green_function}. Over the surface of the sphere, for any polar angle, $\theta$, the electrical potential always will be null. This is equivalent to make the Green function \eqref{full_green_function} vanish for $r=R$, 

\begin{equation}
G(R,\theta,r',\theta')=0,
\end{equation}
what leads to an expression for $G_{\mathsf{L}}$
\begin{equation}\nonumber
G_{\mathsf{L}}(R,\theta,r',\theta')=-\frac{1}{4\pi}\frac{1}{\sqrt{R^2+r'^2-2Rr'\cos(\theta-\theta')}}.
\end{equation}

By inspection, we verify that in the point $r$ the $G_{L}$ must have the following form,

\begin{equation}
G_{\mathsf{L}}(r,\theta,r^{\prime},\theta^{\prime})=-\frac{1}{4\pi}\frac{1}{\sqrt{\frac{r^{\prime\,2}r^2}{R^2}+R^2-2r r^{\prime}\,\cos(\theta-\theta')}},
\label{esfera_green}
\end{equation}

which corresponds to the Green's function inside the sphere, for a point image charge $q^{\prime}$ outside at point $r^{\prime}=\frac{R^{2}}{r^{\prime\,2}}$. The equation \eqref{esfera_green} is the only one that leads to a vanishing Green's function over the surface of the sphere.

Now, we must find the associated electrostatic potential by integrating over the volume in equation \eqref{general_potencial_green}, assuming a charge distribution like $\rho(\vec{r})=\frac{q}{r^{\prime\,2}}\delta(r'-d)\delta(\theta'-0)\delta(\phi'-0)$ which leads to

\begin{equation}
\begin{split}
\varphi(r,\theta)=\frac{1}{4\pi\epsilon_0}&\left\lbrace\frac{q}{\sqrt{r^2+d^2-2rd\cos\theta}}\right.
\\
&\left.-\frac{\left(q R/d\right)}{\sqrt{\left(r^2+\frac{R^{4}}{d^{2}}-2r\frac{R^{2}}{d}\cos\theta\right)}}\right\rbrace.
\end{split}
\label{potencial_esfera_final}
\end{equation}
The result obtained in equation \eqref{potencial_esfera_final} can be derived by using the method of images. Considering a negative image charge placed a distance $r^{\prime}=\frac{R^{2}}{d^{2}}$, from the centre of the spherical shell and a charge $q'=- (q\,R/d)$ produces the same results showed in \eqref{potencial_esfera_final} as represented in the figure \ref{campo_eletrico_esfera_grafico}.

\subsubsection*{Electric Field}

\begin{figure}[htp]
\includegraphics[width=\linewidth]{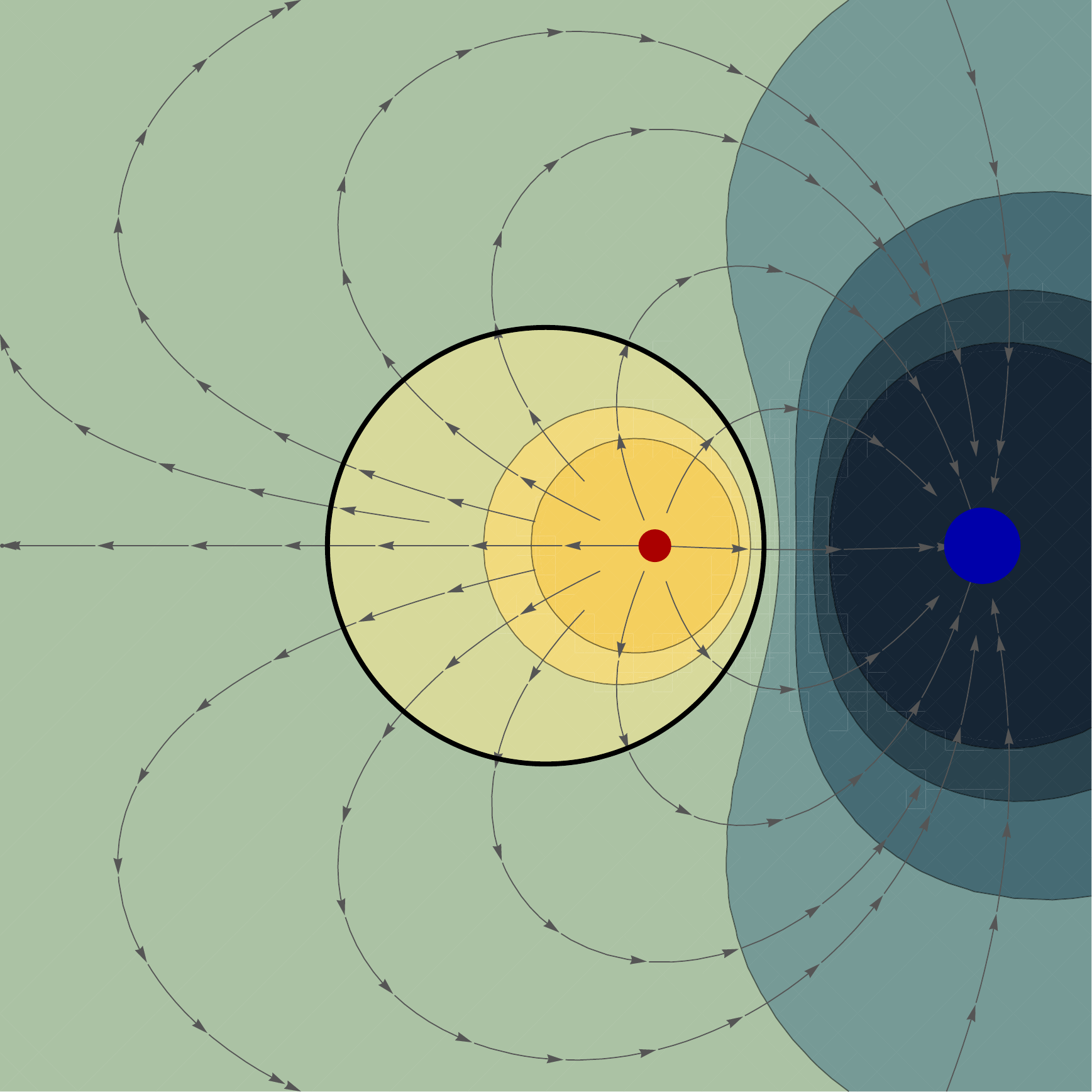}
\caption{Lines of force due to the electrostatic field $\vec{E}(\vec{r})$ and the equipotential surfaces for a positive point charge (red) inside the spherical shell of radius $R$. The blue charge represents the image charge $q^{\prime}$, and it guarantees the electric field is null over the surface of the spherical shell.}     
\label{campo_eletrico_esfera_grafico}
\end{figure}

From the electrostatic potential  \eqref{potencial_esfera_final}, it is possible to find the electric field \eqref{6}, which is expressed in spherical coordinates as 
\begin{equation}
\vec{E}(r,\theta)=E_{r}\,\hat{r}+E_{\theta}\,\hat{\theta},
\end{equation}
where
\begin{equation}
\begin{split}
E_{r}=\frac{q}{4\pi \epsilon_{0}}&\left\lbrace\frac{(r-d\cos\theta)}{\left[r^{2}-2 r d\cos \theta+d^{2}\right]^{3/2}}\right.
\\
&\left.-\left(\frac{R}{d}\right)\frac{\left(r-\frac{R^{2}}{d}\cos\theta\right)}{\left[r^{2}-2 r\frac{R^2}{d}\cos\theta +\frac{R^4}{d^2} \right]^{3/2}}\right\rbrace
\end{split}
\end{equation}
and
\begin{equation}
\begin{split}
E_{\theta}= \frac{q}{4\pi \epsilon_{0}}&\left\lbrace\frac{d \sin \theta}{\left(d^2-2 d r \cos \theta +r^{2}\right)^{3/2}}\right.
\\
&\left.-\frac{R}{d}\frac{\frac{R^2}{d} \sin \theta}{\left(\frac{R^4}{d^2}-\frac{2 r R^{2} \cos \theta}{d}+r^2\right)^{3/2}}\right\rbrace
\end{split}
\end{equation}
and the lines of force are represented in the figure \ref{campo_eletrico_esfera_grafico}.

Considering that the inner charge lies on the z-axis, the induced charge density  at surface of the sphere will be  described by a function of the polar angle $\theta$
\begin{equation}
\begin{split}
\sigma(\theta)&=\epsilon_{0}\left.\frac{\partial V}{\partial r}\right|_{r=R}
\\
&= -\frac{q}{4\pi}\frac{\left(R^2-d^2\right)}{R \left(R^{2} + d^{2}-2 d R \cos \theta\right)^{3/2}},
\end{split}
\end{equation}
and the total charge on the surface of sphere can be found by integrating over all angles, 
\begin{equation}
Q_{t}=\int_{0}^{\pi}\int_{0}^{2\pi}\sigma(\theta)\,d\Omega = -q.
\end{equation}

What would happen if charge $q$ was outside the grounded sphere? In this case, this problem can be solved using this procedure in a similar way. Assuming charge $q$ is located at position $\vec{r}^{\prime}=d$ outside a grounded sphere of radius $R$, the electrostatic potential outside is given by the sum of the potentials due to the charge and its image charge $q^{\prime}$ inside the sphere. 
%
\section{Conclusion}
\label{Conclusion}
It was discussed in this paper the possibility of establishing comparison between Green's function and the method of images in electrostatic problems. The method of images relies upon a strong sense of physical interpretation, while the technique of Green's function is a powerful form of solving problems involving differential equations. Then, it is possible to conclude that Green's function mathematically structure the Method of Images. In other words, this is equivalent to say that mathematics structures physical thought.

The solution attached to the image charge appears as a solution for Laplace's equation, satisfying the boundary conditions associated. On the other hand, Green's function method is more general technique than the one due to calculation by image charges. 

However, in a physics problem, without the interpretation connecting these two instances, mathematical knowledge relates in non-substantive way and may be anchored to non relevant prior knowledge. Therefore, it leads to non elaborated ideas as, for example, `"problems involving conductors are solved by Green's function'' or ``problems involving conductors are solved by the method of images'', what places this kind of relation closer to the rote learning pole and further from the meaningful learning pole \cite{Ausubel}. 

Neverthless, in parallel, the methods may be meaningful both in Physics and Mathematics, once it is possible to learn about conductors in electrostatic equilibrium while conceptually and operationally tackling only using Green's functions.

The authors defend, as does Karam\cite{Karam}, that mathematical knowledge structures physical thought and that gives meaning to mathematical knowledge through situations that make the concept of Green's function useful and meaningful in the field of physics \cite{Vergnaud}, permitting transference to the domain of Ma\-the\-ma\-tics\cite{Rebello}. 

The value of this article underlies in showing a deep relation between physical thought and mathematical structure in a case of electromagnetism (professional cycle). It offers a wider view on the role of Mathematics in Physics than the common views of Mathematics as tool (operationalistic function) or as a merely language (restricted communicative function).

Another intricate point in the discussion is the fact that this knowledge is necessarily tied to epistemological features and these cannot be cast aside in teaching-learning processes. Green himself obviously  did not knew the Dirac Delta function, neither Dirac himself had a formal proof of its validity, which was developed by Schwartz \cite{Schwartz}, but this did not stopped them from doing elaborated Mathematics. 

Similar epistemological difference can be found a\-mong the works of Newton (or Leibniz) and the ones by Weierstrass. For, as much the notion of function due to the latter mathematician approaches the concept of number (static view), the one due to the former in closer to the concept of variable (dynamic view) \cite{Rezende}. 

Related to this, is the unmentionable wide failure in calculus teaching in the first year of any course of exact Sciences \cite{Rezende}, whose cause is, partially, associated with disregarding this feature into teaching-learning processes: students often study textbooks approaching the concept in a Weierstrassian perspective, which is much further (and much more formal) from students' prior knowledge than it should be. In spite of the existence of great teachers and students in theses courses, this epistemological features is beyond their will power or applied didactical methodology in the teaching processes. 

Returning to the discussion of Green's function, we advise that its interpretation should be approached to the notion of point source, as did Green himself, because this can provide conditions for comprehension of more modern concepts as, for example, the Dirac Delta Function. Without this epistemological ingredient, the process of interdisciplinary interaction between Mathematics and Physics in classroom can blatantly fail in reaching its objective of providing conditions for meaningful learning \cite{Ausubel}

The authors expect to contribute, by means of discussion of these simple examples, to demonstrate the feasibility of discussing in an integrated manner the method of images (with high degree of physical interpretation) and the technique of Green's function (with high degree of mathematical power) in classroom. 

The authors also look forward to discuss principles related to providing condition not just for comprehension of the secondary class of situations $\Gamma_E$ pointed in \cite{Author}, but seeking for interdisciplinary integration bet\-ween Physics and Mathematics in a manner of promoting reasoning founded in the thesis that Mathematics structure Physical thought \cite{Karam} and that Physics may give sense to concepts of Mathematics \cite{Vergnaud}.



\renewcommand{\refname}{References}

\end{document}